\documentclass[prd,twocolumn,showpacs,preprintnumbers,amsmath,nofootinbib,amssymb,floatfix]{revtex4}
\usepackage{graphicx}
\usepackage[sort&compress]{natbib}
\usepackage{subfigure}
\usepackage{amsmath}
\usepackage{amsfonts}
\usepackage{cancel}
\usepackage{lmodern,dsfont}
\usepackage{amssymb}
\begin{document}
\arraycolsep1.5pt
\newcommand{\Ima}{\textrm{Im}}
\newcommand{\Rea}{\textrm{Re}}
\newcommand{\mev}{\textrm{ MeV}}
\newcommand{\be}{\begin{equation}}
\newcommand{\ee}{\end{equation}}
\newcommand{\ba}{\begin{eqnarray}}
\newcommand{\ea}{\end{eqnarray}}
\newcommand{\gev}{\textrm{ GeV}}
\newcommand{\nn}{{\nonumber}}
\newcommand{\dtres}{d^{\hspace{0.1mm} 3}\hspace{-0.5mm}}
\newcommand{\rts}{ \sqrt s}
\newcommand{\non}{\nonumber \\[2mm]}

\title{On ambiguities of sign determination of the S-matrix from energy levels in a finite box.}

\author{E. Oset}
\affiliation{
Departamento de F\'{\i}sica Te\'orica and IFIC, Centro Mixto Universidad de Valencia-CSIC,
Institutos de Investigaci\'on de Paterna, Aptdo. 22085, 46071 Valencia, Spain.
}

\date{\today}

\begin{abstract}

  In a recent paper the authors make a study on the determination of the S-matrix elements for scattering of particles in the infinite volume from the energy levels in the finite box for the case of multiple channels. The study is done with a toy model in 1+1 dimension and the authors find that there is some ambiguity in the sign of nondiagonal matrix elements, casting doubts on whether the needed observables in the infinite volume can be obtained from the energy levels of the box. In this paper I present an easy derivation, confirming the ambiguity of the sign and argue that this, however, does not put restrictions in the determination of observables. 

\end{abstract}

\maketitle

\section{Introduction}
\label{Intro}

The determination of scattering amplitudes and hadron spectra is one of the challenging tasks
of Lattice QCD and many efforts are being devoted to this problem
\cite{Nakahara:1999vy,Mathur:2006bs,Basak:2007kj,Bulava:2010yg,Morningstar:2010ae,Foley:2010te,Alford:2000mm,Kunihiro:2003yj,Suganuma:2005ds,Hart:2006ps,Wada:2007cp,Prelovsek:2010gm,Lin:2008pr,Gattringer:2008vj,Engel:2010my,Mahbub:2010me,Edwards:2011jj,Lang:2011mn,Prelovsek:2011im,Melnitchouk,Nemoto,fxlee,TBurch,ishii, TTTakahashi,mena}.  For the case of one channel scattering, and resonances decaying in just one channels,  L\"uscher's approach is often used \cite{luscher,Luscher:1990ux}. The method allows to reproduce the phase shifts from the discrete energy levels in the box. This method has been recently simplified and improved in
Ref.~\cite{misha} by keeping the full relativistic two body propagator
(L\"uscher's approach keeps the imaginary part of this propagator exactly but makes approximations on the real part) and extending the method to two or
more coupled channels. The method has also been applied in
Ref.~\cite{mishajuelich} to obtain finite volume results from the J\"ulich model for meson baryon interaction and in Ref.~\cite{albertolam} to determine the strategy to find the two poles of the $\Lambda(1405)$ from lattice QCD simulations. Further applications and references to works done along these lines can be seen in \cite{murcia}. 

The extension of  L\"uscher's approach to coupled channels has been addressed in several works \cite{Lage:2009zv,misha,Polejaeva:2012ut,Hansen:2012tf,Bernard:2012bi}.
 In the work of Ref.~\cite{misha}, the inverse problem of getting phase shifts and resonances from lattice results using two channels was addressed, paying special attention to the evaluation of errors and the precision needed on the lattice results to obtain phase shifts and resonance properties with a desired accuracy. 
  
  In a recent paper \cite{cohen} a coupled channel study of this inverse problem is done with two channels, using a toy model in 1+1 dimension and it is concluded that the determination of the S-matrix in the case of T-invariance (we only consider this case) has an ambiguity in the sign of the nondiagonal matrix element. Due to this, doubts are cast that in a realistic case this does not pose problems in the determination of needed observables. 
  
  In the present paper we present a derivation of this inverse problem in the realistic case and conclude that indeed this ambiguity arises. However, we also argue that this ambiguity, tied to an arbitrary phase in the wave function of one channel with respect to the other, has no repercussion in the determination of observables.

\section{Formalism}

 In the chiral unitary approach the scattering matrix in coupled
channels is given by the Bethe-Salpeter equation in its factorized form \cite{npa}. We assume just s-waves for simplicity (see generalization to other waves in \cite{mixing}) and we have

\be
T=[1-VG]^{-1}V= [V^{-1}-G]^{-1},
\label{bse}
\ee
where $V$ is the matrix for the transition potentials between the
channels and 
$G$ 
is a diagonal matrix with the $i^{\rm th}$
element, $G_i$,
given by the loop function of two propagators (we shall use two mesons),
which is defined as 
\be
\label{loop}
G_i=i \,\int\frac{d^4 p}{(2\pi)^4} \,
\frac{1}{(P-p)^2-M_i^2+i\epsilon}\,\frac{1}{p^2-m_i^2+i\epsilon}
\ ,
\ee
where $m_i$ and $M_i$ are the masses of the pair of mesons  
and $P$ the four-momentum of the global meson-baryon system.

The loop function in Eq.~(\ref{loop}) needs to be regularized and
this can be accomplished either with dimensional regularization
or with a three-momentum
cutoff. The equivalence
 of both methods was
shown in Refs.~\cite{ollerulf,ramonetiam}.

In the cut off method a cutoff in three momentum is used 
once the $p^0$ integration is analytically performed \cite{npa}, and one
gets

\ba
&&G_i=\hspace{-4mm}\int\limits_{|\vec p|<p_{\rm max}}
\frac{d^3\vec p}{(2\pi)^3}\frac{1}{2\omega_1(\vec p)\,\omega_2(\vec p)}
\frac{\omega_1(\vec p)+\omega_2(\vec p)}
{E^2-(\omega_1(\vec p)+\omega_2(\vec p))^2+i\epsilon},
\non 
&&\omega_{1,2}(\vec p)=\sqrt{m_{1,2}^2+\vec p^{\,\,2}}\, ,
\label{prop_cont}
\ea
with $m_1$, $m_2$ corresponding to $m_i$ and $M_i$ of Eq.~(\ref{loop}).

When one wants to obtain the energy levels in the finite box,
 instead of integrating over the
energy states of the continuum, with $p$ being a continuous variable
as in Eq.~(\ref{prop_cont}), one must sum over 
the discrete momenta allowed
in a finite box of side $L$ with periodic boundary conditions.
We then have to replace $G$ by 
$\widetilde G={\rm diag}\,(\widetilde G_1,\widetilde G_2)$ (in two channels), where 
\ba
\widetilde G_{j}&=&\frac{1}{L^3}\sum_{\vec p}^{|\vec p|<p_{\rm max}}
\frac{1}{2\omega_1(\vec p)\,\omega_2(\vec p)}\,\,
\frac{\omega_1(\vec p)+\omega_2(\vec p)}
{E^2-(\omega_1(\vec p)+\omega_2(\vec p))^2},
\non 
\vec p&=&\frac{2\pi}{L}\,\vec n,
\quad\vec n\in \mathds{Z}^3 \,
\label{tildeg}
\ea

 This is the procedure followed in Ref.~\cite{misha}.  The eigenenergies of
the box correspond to energies  that produce poles in the $T$ matrix, 
Eq.~(\ref{bse}), which in the finite box correspond to zeros of the
determinant of $1-V\widetilde G$,
 \be
\label{eq:det}
\det(1-V\tilde G)=0\, .
\ee  
For the case of two coupled channels Eq.~(\ref{eq:det}) can be written as
\begin{align}
\det(1-V\tilde G)&=1-V_{11}\tilde G_1-V_{22}\tilde G_2\nonumber\\
&\quad+(V_{11}V_{22}-V_{12}^2)\tilde G_1\tilde G_2\nonumber\\
&=0\,.\label{det2}
\end{align} 

One can already see there that $V_{12}$ appears squared, hence, a change of sign in it will not change the spectra of levels in the box. We shall see that this is also the case in the inverse problem and we can only determine $T_{12}^2$. However, this will not prevent us from determining the three scattering magnitudes in this case, the two phase shifts and the inelasticity.

\section{The inverse problem}

The inverse problem of obtaining $T_{ij}$ from the energy levels in the box is most efficiently written in terms of $\delta G$ defined as
\be 
\delta G\equiv \tilde{G}-G, 
\ee
the magnitude used in \cite{arXiv:1108.5371}, which is finite in the limit of $p_{max} \to \infty$, the limit taken in \cite{arXiv:1108.5371,albertolam}, and proportional to the  L\"uscher function \cite{Beane:2003yx,misha}.

 Let us start from Eq.~(\ref{bse}) that gives the $T$ matrix in the infinite volume and write the correspondent  scattering matrix in the finite volume, $\tilde{T}$:

\begin{align}
\tilde{T}=[V^{-1}-\tilde{G}]^{-1}. \label{a}
\end{align}

Note that for the case of one channel the poles of $\tilde{G}$ provide $V^{-1}-\tilde{G}= 0$ and, thus, $V^{-1}=\tilde{G}$ for the eigenenergies of the box. Then we can recast Eq. (\ref{bse}) as

\begin{equation}
T(E)=[\tilde{G}(E)-G(E)]^{-1}.\label{tluescher}
\end{equation}
and this is the formulation of L\"uscher's formula in \cite{misha}.

Coming back to the multichannel problem and
using Eqs.~(\ref{bse}) and (\ref{a}), as done in \cite{albertolam}, we get
\begin{align}
\tilde{T}^{-1}&=T^{-1}-\delta G=T^{-1}[1-T\,\delta G], \label{b}
\end{align}
 Hence,
\begin{align}
\tilde{T}=[1-T\,\delta G]^{-1} T. \label{c}
\end{align}
which allows us to get $\tilde{T}$ directly in term of $T$, without going through an intermediate potential.
One can note that this formula is like the one of Eq.~(\ref{bse}), or Eq.~(\ref{a}) for $\tilde{T}$, substituting $V\to T$ and $\tilde{G}\to \delta G$. Hence,
the condition to obtain the energy levels in the box, $\textrm{det}(\tilde{T})=0$, leads to the analogous secular equation of Eq.~(\ref{det2}) in terms of $T$ and $\delta G$ substituting $V$ and $\tilde{G}$, respectively,
\begin{align}
(1-T_{11}\,\delta G_{11})(1-T_{22} \delta G_2)-T^2_{12}\delta G_1 \delta G_2=0,
\end{align}
or equivalently,

\begin{align}
T_{11}\,\delta G_{11}+T_{22} \delta G_2-(T_{11}T_{22}-T^2_{12})\delta G_1 \delta G_2=1.
\label{secular}
\end{align}

It is clear that with just one energy eigenvalue for a given L, Eq. (\ref{secular}) cannot provide the full $T_{ij}$ matrix. In \cite{misha} different methods were used to get $T$ from the eigenenergies of the box. The simplest conceptually is to take a certain energy from three levels, which correspond to three different values of L, and then determine the three values of  $T_{ij}$. Actually, what one determines is $T_{11}$, $T_{22}$ and $T^2_{12}$. So, we can see in a realistic case that we only obtain $T^2_{12}$. However, this is not a problem to determine the observables. This indetermination should be there because a change of relative sign of the wave function for the states 1 and 2 leaves $T_{11}$, $T_{22}$ unchanged but it changes the sign of $T_{12}$, and the physics cannot depend on this sign. To make this more explicit we write explicitly the S-matrix in terms of the observables in the next subsection.

\subsection{Phase shifts and Inelasticities:}

In order to obtain the phase shifts and
inelasticities we use the two-channel $S$ matrix \cite{Weinstein:1990gu}

\begin{equation}
S = \left[ \begin{array}{ll}
\eta e^{2 i \delta_1} & i (1 - \eta^2)^{1/2} \,  e^{i (\delta_1 + \delta_2)}\\
i (1 - \eta^2)^{1/2} e^{i (\delta_1 + \delta_2)} & \eta e^{2 i \delta_2}
\end{array} \right]
\label{smatrix}
\end{equation}

\noindent
where $\delta_1, \delta_2$ are the phase shifts for the 1 and 2 channels
and $\eta$ is the inelasticity. The elements in the $S$ matrix
are related to our amplitudes via:

\begin{equation}
\begin{array}{l}
T_{11} = - \frac{8 \pi \sqrt{s}}{2 i p_1} (S_{11} - 1)\\[2ex]
T_{22} = - \frac{8 \pi \sqrt{s}}{2 i p_2} (S_{22} - 1)\\[2ex]
T_{12} =  T_{21} = - \frac{8 \pi \sqrt{s}}{2 i \sqrt{p_1 p_2}} S_{12}\\[2ex]
\end{array}
\end{equation}

It is interesting to recall that the first two equations allow us to 
determine $\delta_1, \delta_2$ and $\eta$ while the third equation allows
us to determine $\delta_1+\delta_2$  and  $\eta$.
 The S-matrix should be unitary and one can see that this is the case even if we change the sign of $S_{12}$. In fact the choice of the positive sign in the square root of $(1 - \eta^2)$ is a matter of convention. One may argue that if we change the sign of $S_{12}$ and use 
Eq. (\ref{smatrix}), one would obtain $e^{i(\delta_1 + \delta_2)}$ with opposite sign to what one would get from using the same matrix element before changing the sign. This reflects the arbitrariness of $\pi$ in the phase shifts. In view of this arbitrariness in the sign, the quantities that should be used are $|T_{12}|^2$, which determines $(1- \eta^2)$ and then $T^2_{12}$ which determines $(1 - \eta^2)e^{2i (\delta_1 + \delta_2)}$ and allows one to get $\delta_1 + \delta_2$, independently of the sign of $T_{12}$. 

\section{Conclusions}

 In this paper we have done a derivation of the inverse method to get the scattering matrix from the energy levels of the system in a finite box. We observe, in agreement with the findings with the toy model of \cite{cohen}, that for the case of two channels studied in \cite{cohen}, the sign of the off diagonal matrix element $T_{12}$ is not defined. Yet, we could see that this had not repercussion in the determination of the observables $\delta_1$, $\delta_2$ and $\eta$, although given the ambiguity on this sign, tied to an arbitrary relative phase in the wave functions of the states, a particular algorithm must be taken to determine the phase shifts from $T_{12}$.
 
 The two channel system that we had in mind was the one of two physical states that couple, say $\pi \pi$ and $K \bar K$, studied for instance in \cite{misha}.
 One may think of other coupled channels systems, like one physical state with two coupled partial waves, where signs and interference are important in angular distributions (the authors of \cite{cohen} might have such and idea going beyond the toy model used). An example of this is the deuteron or dineutron system in the presence of a tensor force \cite{newassum}. The issue of partial wave mixing is an interesting one in finite volume because a square box breaks rotational invariance and this leads invariably to partial wave mixing. So, the problem is well documented \cite{Luscher:1990ux}. Closer to the case of explicit L mixing caused by particular external forces is the case of the L mixing in the moving frame. Here one still has a central potential but the imposition of the boundary conditions in a frame where the total momentum of the system is not zero leads to partial wave mixing. This problem has also received much attention   
\cite{mixing,Kreuzer:2012sr,Rummukainen:1995vs,Kim:2005gf,Bour:2011ef,Beane:2011sc,Davoudi:2011md,Fu:2011xz,Leskovec:2012gb,Dudek:2012gj,Hansen:2012tf,Briceno:2012yi,akakinew}. The exposition and solution of the problem is made in a relatively simple and pedagogical way in \cite{mixing}, where L mixing and physical coupled channels are considered simultaneously. There one can see that the equations and strategies that allow one to obtain the different partial ways and their relative signs are far more involved than the simple equations used in the present work or in \cite{cohen}. The result of these works is that the inverse problem has solution without ambiguities and realistic cases are even solved explicitly in \cite{mixing}. 

     Whether in some finite volume topic ambiguities appear in the inverse problem is an open issue. What is clear is that the ambiguity found in the present work dealing with just one partial wave, or in \cite{cohen} in the 1+1 toy model, where partial waves cannot be defined, poses no problem for any of the works done with finite volume so far. 

\section*{Acknowledgments}
 The author would like to thank T. Cohen and A. Parre\~no for discussions and Michael D\"oring for a careful reading of the manuscript. This work is partly supported by DGICYT contract FIS2011-28853-C02-01, FEDER funds of the EU, the Generalitat Valenciana in the program Prometeo, 2009/090, and the EU Integrated Infrastructure Initiative Hadron Physics 3
Project under Grant Agreement no. 283286.

\end{document}